\documentclass[10pt,a4paper,final,notitlepage,twocolumn]{iopart}
\usepackage{graphicx}% Include figure files
\usepackage{dcolumn}% Align table columns on decimal point
\usepackage{bm}% bold math
\usepackage{iopams}% bold math

\begin{document}

\title[Ab initio correlation effects on MPc-based devices]{Ab initio correlation
effects on the electronic and transport properties of
metal(II)-phthalocyanine based devices}

\author{Arrigo Calzolari$^{1}$, Andrea Ferretti$^{1,2}$,
        Marco Buongiorno Nardelli$^{3}$}

\address{$^1$ INFM-CNR-S3 National Center on
              nanoStructures and bioSystems at Surfaces, I--41100 Modena, Italy.
}
\address{$^2$ Dipartimento di Fisica, Universit\`a di Modena e Reggio Emilia,
              I--41100 Modena, Italy.
}
\address{$^3$ CHiPS and Department of Physics, NCSU, Raleigh, NC
              27695-7518 and CSMD, Oak Ridge National Laboratory, Oak Ridge, TN
              37831-6359.}

\ead{calzolari.arrigo@unimore.it}
\date{\today}

%*******************************

%  ABSTRACT

   %*******************************

\begin{abstract}

Using first principles calculations  in the framework of Density
Functional Theory,  we investigated the electronic and transport
properties of metal(II)-phthalocyanine (M(II)Pc) systems, both in a
single molecule configuration and in a model-device geometry. In
particular, using the Copper(II)- and Manganese(II)-Pc as
prototypical examples, we studied how electronic correlations on the
central metal-ion influence the analysis of the electronic
structure of the system and we demonstrated that the choice of the
exchange-correlation functional, also beyond the standard local or
gradient corrected level, is of crucial importance for a correct
interpretation of the data. Finally, our electronic transport simulations have shown that
M(II)Pc-based devices can act selectively as molecular conductors,
as in the case of Copper, or as spin valves, as in the case of
Manganese, demonstrating once more the great potential of these
systems for molecular nanoelectronics applications.
\end{abstract}

\pacs{  }
%\maketitle

%*******************************

%  INTRO

%*******************************

\section{Introduction}\label{intro}

\begin{figure}
\begin{center}
\includegraphics[width=0.90\textwidth]{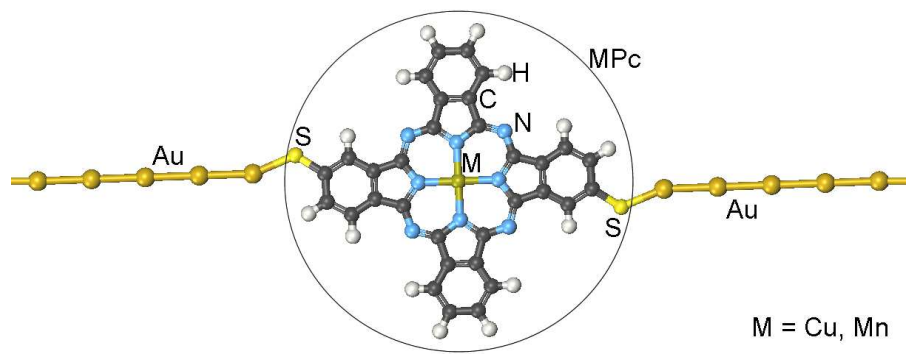}
\caption{(color online) Atomic structure of a generic
metal(II)-phthalocyanine (MPc) bridged between two one-dimensional
gold chains in a lead/conductor/lead geometry. The circle marks
the phthalocyanine molecule. Different colors identify different
chemical species: grey=C, white=H, cyan=N, gold=Au, yellow=S,
green=M; M stands for either Copper or Manganese. In the single
molecule configuration the Au chains are absent and the two S
atoms are replaced by H atoms.}\label{figure1}
\end{center}
\end{figure}

Metal(II)-phthalocyanines (MPc) have attracted a growing interest
in view of their potential for nanoscale applications, such as
light-emitting diodes, field effect transistors, photovoltaic
cells and single molecule devices~\cite{kadi+03book}. In
particular, metal(II)-phthalocyanines are considered prototypical
organic semiconductor materials, and their properties have been
widely investigated in a variety of experimental conditions. The
characteristics of the MPc/substrate interactions
~\cite{aris+05prb,lei+06jpcb,coss+04jpcb,suzu+02jpcb}, the effects
of doping ~\cite{crac+05jacs,ding-gao06ass,gao-yan03cpl}, and the
adsorption of additional small molecules~\cite{szcz-brag95vac}
have been investigated with techniques such as photoemission
spectroscopy~\cite{down+04cpl,zurm+01prb,lozz+06jvsta,schw+02prb},
scanning tunneling
spectroscopy~\cite{nazi+03sci,nazi+05pnas,mika+06jcp} and
microscopy~\cite{lipp+89prl,lu+96jacs,hipp+96jpc}, electronic
paramagnetic resonance~\cite{fina+06pccp}, NMR~\cite{fili+07prb},
gas-phase electron diffraction~\cite{mast+00jms}, etc. However, it
is not unfrequent that different experiments produce   somewhat
contradicting results
~\cite{crac+05jacs,down+04cpl,zurm+01prb,giov+07jcp,molo+05jap,schw+03apl}
and a careful interpretation of these data in terms of a robust
theoretical analysis has become mandatory.

An accurate theoretical determination of the electronic structure
of the single molecule is the first step towards a meaningful
interpretation of these complex systems: understanding the
sequence of the frontier molecular orbitals along with their spin
order is of paramount importance for the analysis of
photoemission, optical and transport experiments.

Phthalocyanines (Pc's) are planar organic macrocycles that can
host a central metal ion (typically a $3d$ transition metal), as
shown in Figure~\ref{figure1}. Because of the relatively large
size of the molecule, the use of {\em ab initio}
approaches~\cite{day+98jmst,rosa-baer94ic,liao-sche01jcp,lozz+04jcp,nguy-pach01jcp,bial+03tsf,tada+05jpcb,liao+05ic}
to investigate the complex metal-ligand coupling has become viable
only recently. However, different electronic structures have been
already proposed, as in the case of the Copper(II)-Pc (CuPc), for
which a partial molecular orbital inversion not observed in
experiments has been
reported~\cite{day+98jmst,rosa-baer94ic,tada+05jpcb}.
Discrepancies like this can be attributed mostly to the level of
description of the electronic correlations in the simulations. On
the basis of the electron counting rules for the Pc ligands the
central ion has a  formal $+2$ charge. Therefore, the presence of
open shell $d$-metals requires a highly accurate description of
the electron-electron interactions and of the spin-ordering in the
MPc.

In order to understand and resolve the ambiguities in the
assignment of the electronic levels, we have
studied how different descriptions of the correlation modify the
electronic structure of these systems. Using the Copper(II)- and
Manganese(II)-Pc as prototypical examples, we have demonstrated
that the choice of the exchange-correlation (XC) functional beyond
the standard local or gradient corrected level is of crucial
importance for a correct interpretation of the data. Moreover, we
have analyzed the effects of the electron correlation on the
electronic transport through a model device: a M(II)Pc molecule
bridged between two atomic chains of gold atoms in a typical
lead/conductor/lead configuration, as shown in Fig.~\ref{figure1}.
This system has been recently proposed as a prototypical
one-dimensional molecular device ~\cite{nazi+03sci,tada+05jpcb}.

%*******************************

%  METHOD

%*******************************

\section{Computational methodology}\label{method}

Our studies are based on state-of-the-art {\em ab initio}  Density
Functional Theory calculations, as implemented in the
\textsc{PWscf} code~\cite{Pwscf}. At the standard level, the
exchange and correlation functional is evaluated within the Local
Spin Density Approximation (LSDA) \cite{perd-zung81prb} and the
Generalized Gradient Approximation (GGA) as proposed in
Ref.~\cite{perd+96prl} (PBE), always including spin polarization
(unrestricted calculations). The electron-ion interaction is
described by {\em ab initio} ultrasoft pseudopotentials
\cite{vand90prb}. For Manganese, we have included the 3$s$3$p$
semicore shells in the valence. The electronic wavefunctions
(densities) were expanded on a plane-wave basis set up to a
kinetic energy cutoff of 25 Ry (200 Ry). The effects of the
electron-electron correlation beyond LDA/GGA have been included
using the DFT+U scheme, in the linear response approach proposed
by Cococcioni and De Gironcoli in 2005 \cite{coco-degi05prb}.

In all the calculations we have used supercells of dimensions
44.94 $\times$ 26.00 $\times$ 12.00 \AA$^3$ with at least 12 \AA\
of vacuum space between adjacent replica in each direction. The
metal/molecule/metal bridge (labeled (Au$_5$S)$_2$M-Pc) was
modeled via one dimensional gold chains connected to the molecule
through sulfur atoms (see Fig. 1). All the structures have been
thoroughly relaxed.

The transport properties of the system have been computed in a
lead/molecule/lead geometry using the fully first principles
approach implemented in the \textsc{WanT}
code~\cite{WanT,calz+04prb}. The method combines an accurate
description of the electronic ground state, provided by {\em ab
initio} DFT calculations, with the Landauer approach to describe
transport properties of extended
systems~\cite{land70pm,datt95book}. The connection is realized by
transforming the Bloch orbitals into maximally-localized Wannier
functions~\cite{marz-vand97prb,ferr+07jpcm}. This representation
naturally introduces the ground-state electronic structure into
the real-space Green's function scheme, which is our tool for the
evaluation of the Landauer quantum conductance.

The \textsc{WanT} method has been also extended in order to
include short range electron-electron (e--e) interactions on
transport properties~\cite{ferr+05prl,ferr+05prb,dara+07prb}. The
effect of the e--e coupling is taken into account by including a
further correlation self-energy term in the calculation of the
conductor Green's function. The e--e self-energy is computed here
using the Three-Body Scattering formalism
(3BS)~\cite{cala-mang94prb,ThreeBS}, which relies on an effective
generalized Hubbard Hamiltonian, and it is solved in a
configuration interaction scheme where up to three bodies are
added to the non-interacting Fermi sea.

\section{Results and Discussions}\label{results}

%*******************************

\begin{figure}
\begin{center}
\includegraphics[width=1.00\textwidth]{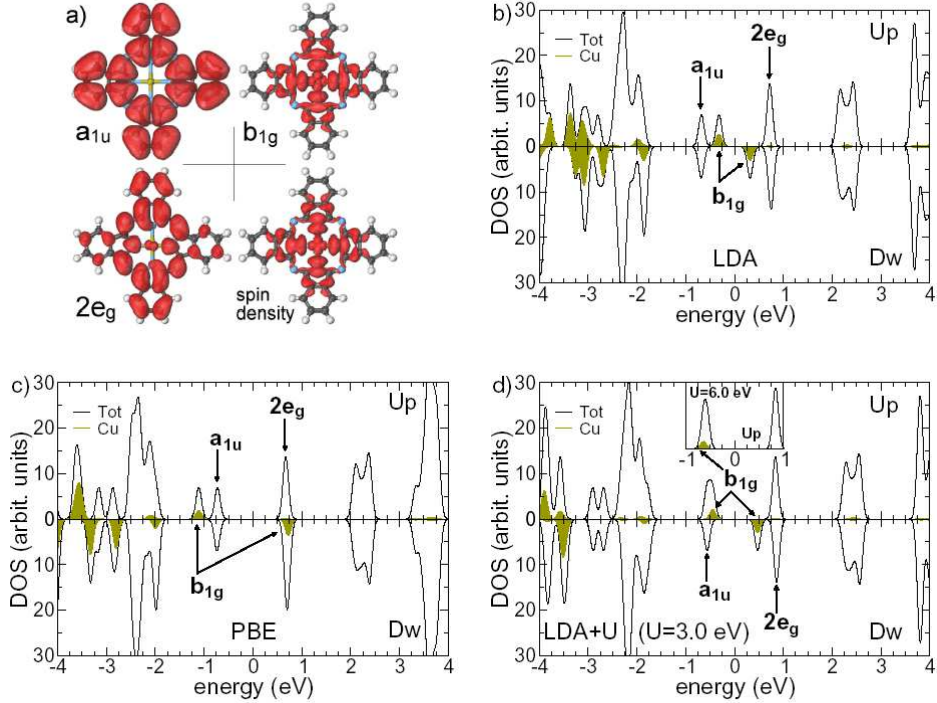}
\caption{(color online) CuPc electronic structure. (a) Charge
density isosurfaces of selected molecular orbitals and spin
density ($\rho_{up}-\rho_{dw}$) for CuPc. Spin-resolved density of
states (DOS) corresponding to (b) LDA, (c) PBE, (d) LDA+U (U= 3eV)
functionals. Shaded green areas  are the projections on the atomic
Cu states. Vertical arrows and labels identify the energy position
of the peaks corresponding to the selected molecular orbitals,
reported in panel (a). Inset of panel (d) shows the spin-up
frontier peaks for the case U= 6.0 eV. The zero energy
discriminates  the occupied and the empty states for each system.
}\label{figure2}
\end{center}
\end{figure}

%*******************************
%  CuPc
%*******************************

\subsection{Copper(II)-phtalocyanine systems}\label{cupc_mol}
\subsubsection{Single molecule.}
The single particle orbitals of the planar CuPc molecules can be
classified according to the irreducible representation of the
$D_{4h}$ group. In this representation the $3d$ orbitals split
into $a_{1g}$ ($d_{z^2}$), $b_{1g}$ ($d_{x^2-y^2}$), $e_g$
($d_{xz},d_{yz}$) and $b_{2g}$ ($d_{xy}$). One of the open
problems in the interpretation of the experimental data is the
energy position of the $b_{1g}$ orbital  with respect to the
$a_{1u}$ ($\pi$) and $2e_{g}$ ($\pi^*$) states, which are the
highest occupied (HOMO)  and the lowest unoccupied molecular
orbital (LUMO) of the Pc rings respectively. The $b_{1g}$ orbital
has a $\sigma$ character and is localized on the copper ion,
whereas the $a_{1u}$ and $2e_{g}$ are single and double
degenerate $\pi$-like states, delocalized over the entire Pc
macrocycles [see Fig.~\ref{figure2}(a)].

In order to clarify the issue, we have calculated the electronic
structure of the isolated CuPc molecule by using two different XC
correlation functionals, namely LDA and PBE. The resulting density
of states are reported in Fig.~\ref{figure2}(b-c). The analysis of
the electronic structure shows the presence of mixed metal-ligand
states in the low-energy part of the spectrum, especially between
the copper ion and the pirole rings, the fingerprint of a direct
interaction between the two subsystems. In Fig.~\ref{figure2}, we
show the $b_{1g}$, $a_{1u}$ and $2e_{g}$ states that are the
frontier orbitals of the overall CuPc molecule. In both LDA and
PBE case, the $a_{1u}$ and the $2e_{g}$ states are spin-degenerate
({\em i.e.} the same energy for both spin components), and the
corresponding energy differences $\Delta$E=E($2e_{g}$)-E($a_{1u}$)
are very similar: 1.40 eV and 1.42 eV for LDA and PBE
respectively. Independently from the choice of the XC functional,
the CuPc molecule has a resulting total magnetization of 1.0
$\mu_B$, due only to the unpaired $b_{1g}$ state, as shown in the
plot of the spin density [Fig.~\ref{figure2}(a)]. However, the
relative order of the states is different: in LDA
[Fig.~\ref{figure2}(b)], the $b_{1g}$ orbital is the
single-occupied HOMO (SOMO) of the overall CuPc molecule, in
agreement with the results by Liao et al.~\cite{liao-sche01jcp};
on the contrary, in PBE the $b_{1g}$ is the HOMO-1 state for the
spin-up part, while it is almost degenerate with $2e_{g}$ orbital
for the spin-down component. The experimental photoemission
results \cite{down+04cpl} seem to indicate that the LDA
configuration is the correct one; however, due to the small
cross-section of the metallic states, no truly conclusive
statements can be drawn \cite{lozz+04jcp}. It is worth noticing
that this result is not biased by the choice of the basis set used
in the calculations. We calculated the electronic structure of the
molecule also using a localized atom-centered numeric basis set
using the \textsc{Dmol}$^3$ code~\cite{dell00cms}: with both LDA
and PBE XC-functionals we obtained the same orbital inversion
observed before. The discrepancies between LDA and PBE highlight
the delicate balance occurring in the electronic structure of this
molecule.

In our DFT calculations the net charge transfer from Cu atom to
the rest of the environment is symmetric and smaller than the
formal one: the analysis of the L\"owdin charges indicates that
the copper atom is positively charged $\delta$q(Cu)= +1.02 e, and
the $3d$ shell hosts $\sim$9.4 electrons. Even though the
calculated values are smaller than the formal ones, the
correlation effects, due to the opening of the $3d$ shell must be
taken carefully into account, especially if one attempts to use
the ground state structure as the basis for the interpretation of
effects (such as doping) that imply the variation of the
oxidation/reduction state of the central ion.

In order to gain further insights on this issue, we introduced
electronic correlation effects beyond standard DFT by including a
local Hubbard U term  on the Cu ion. We tested several values of
the U parameter (U=0.5, 2.0, 3.0, 4.0, 6.0 eV); the results for
U=3 eV are shown in Fig.~\ref{figure2}(d). The Hubbard term
operates mainly on the energy of the ion-centered states, leaving
the Pc ones unchanged. We note a downward (upward) shift of the
$b_{1g}$ state for the spin-up (spin-down) component,
respectively. The net effect is the opening of the SOMO-LUMO gap.
By increasing the value of the Hubbard U, the shift of the
$b_{1g}$ peak increases accordingly. In particular, for U=0.5 and
2.0 eV, both DOS resemble the LDA spectra of
Fig.~\ref{figure2}(b). For U=3.0 eV, even thought $b_{1g}$ is
still the SOMO, it is almost degenerate with the spin-up $a_{1u}$
state [Fig. \ref{figure2}(d)]. This effective degeneracy is
observed also for the other two values of the Hubbard parameter
(U=4.0, 6.0 eV) that we considered.

By further increasing U, we observe a slight shift of the $b_{1g}$
peak, that leads to an inversion of the $b_{1g}$ and $a_{1u}$
states for U= 6.0 eV [see inset of Fig.~\ref{figure2}(d)]. This
trend confirms how the order of the single particle states is
strongly related to the specific treatment of the
electron-electron correlation. However, an on-site correlation
parameter U=6.0 eV is a rather high value, greater than any of
those  experimentally measured for similar systems
\cite{note_hubbard}. Thus, it is reasonable to conclude that the
LDA order of the states is the correct one, while the exact
position of the $b_{1g}$ peak wrt the Pc's states depends on the
correct value of the electron correlation. From our calculations
it results also that in the case of CuPc the PBE XC-functional
overestimates the e-e correlation, as if a strong Hubbard
potential were applied \cite{note_hubbard1}.

\subsubsection{Device configuration.}

%******************************
\begin{figure}
\begin{center}
\includegraphics[width=1.0\textwidth]{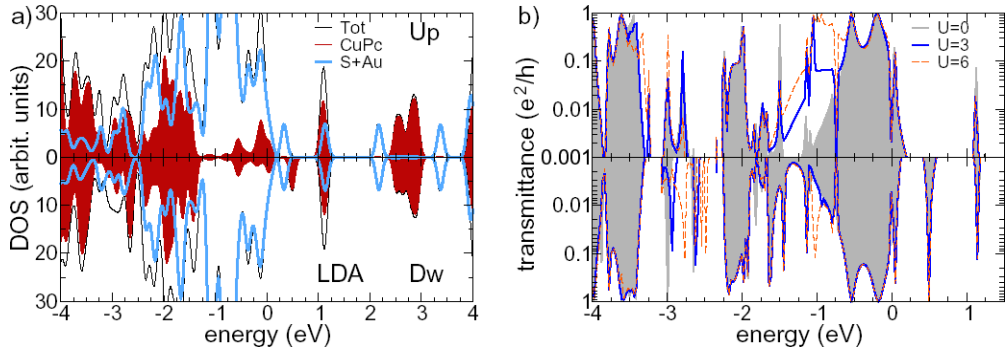}
\caption{(color online) (a) Spin-polarized DOS of the periodic
(Au$_5$S)$_2$CuPc system. Shaded red area (thick blue line) is the
projection on the CuPc (S-Au) atomic states. (b) Mean field
(shaded area) and correlated (straight and dashed lines)
two-terminal transmittance of open CuPc-based device,
spin-resolved. The zero energy references are set to Fermi level
of the periodic (Au$_5$S)$_2$CuPc system.}\label{figure3}
\end{center}
\end{figure}

%*******************************

The system we studied is inspired by a prototypical
one-dimensional single-molecule device originally proposed by
Nazin et al. \cite{nazi+03sci}. The original system was realized
placing a CuPc molecule between two short chains of gold atoms,
deposited on a NiAl substrate. The electronic structure was then
characterized by STS measurements. The metallic character of the
substrate prevented the direct measurement of I-V characteristics
through the double interface. The coherent transport properties of
the open device have been calculated so far by Tada and coworkers
\cite{tada+05jpcb}, who found a quantum conductance of the order
of $G=3\times10^{-4}$ e$^2/h$. This value, smaller than the
typical ones ($G=0.01$ e$^2/h$) observed for aromatic
molecules~\cite{xu-tao03sci}, is mostly due to the weak gold/CuPc
interaction, {\it i.e} to the absence of a direct chemical bond.
For these reasons we slightly modified the geometry of the system,
connecting the molecule and the leads, through a sulfur bridge
(Fig.~\ref{figure1}). The choice of the sulfur functionalization
is justified by the well-known ability of thiol groups to anchor
molecules to gold electrodes~\cite{schr00pss}.

Following the method~\cite{calz+04prb} described in
Section~\ref{method}, we have simulated an open device where the
Au wires constitute the semi-infinite leads and the CuPc acts as
the conductor. On the basis of the results for a single molecule,
we calculated the ground state electronic structure using LDA
XC-functional. The spin-resolved electronic structure of the
periodic (Au$_5$S)$_2$CuPc building block [Fig.~\ref{figure3}(a)]
is characterized by a finite density of states at the Fermi level
(E$_F$=0 eV): while the formation of the nanojunction induces a
realignment of the molecular and gold orbitals, the CuPc molecule
perturbs the pure one-dimensional features of the infinite Au
chain, such as the van Hove singularities, and the dispersive
$sd$-band crossing the Fermi level
\cite{calz+04prl,calz-buon05prb}. As expected, the presence of the
thiol bridge at the metal/molecule interface favors the mixing of
the molecular orbitals. Indeed, the overall SOMO for the spin-up
component is the result of the hybridization of the delocalized
$sd$-like states of the pristine chain and the $b_{1g}$ state of
the CuPc molecule. Other hybridized orbitals, which involve also
the $a_{1u}$ state of the molecule, are recognizable in the energy
range from $E=-0.5$ eV  to Fermi level. The (Au$_5$S)$_2$CuPc
system maintains a global magnetic polarization of 1.0 $\mu_B$,
due to the single occupation of the $b_{1g}$ state derived from
the isolated molecule. However, the effect of this unbalance  on
the overall transmission is reduced by the spin-unpolarized
density of states of the gold wire.

As a starting point, we calculated the electronic transport at the
mean field level: we treated the electron correlation within DFT
(LDA), by using the Landauer formula~\cite{datt95book}. In
particular, we focused on the zero bias regime; in this case the
quantity to be calculated is the quantum transmittance $T(E)$. The
value of the transmittance at the Fermi level gives the quantum
conductance.

The results for the spin-up and spin-down components of the
quantum transmittance (logarithmic scale) are shown in Fig.~%
\ref{figure3}(b). The corresponding quantum conductance are
$G_{up}=0.059$ e$^2/h$ and $G_{dw}=0.064$ e$^2/h$. These values
are reasonably high,  when compared  with those calculated for the
unbound system~\cite{tada+05jpcb} or with the gold chain
\cite{agra+03prep} (which gives almost $G=1.0$ 2e$^2/h$). This is
due to the important mixing of the CuPc orbitals with the
$sd$-bands of the gold wire. In other energy regions, such as in
the range $E\in [-2.0, -0.5]$ eV, that corresponds to the $5d$
band of the Au wire, the orbital mixing is very small, leading to
a localization of the electronic states and a corresponding
reduction of the transmittance.

As a further test, we calculated the transport properties starting
from PBE-DFT calculations. In that case the inversion of the CuPc
states near the Fermi level is almost irrelevant, due to the same
coupling with the unpolarized $sd$ manyfold of the gold wire.

In order to take into account the effect of the electron
correlation, we evaluated the additional e-e self-energy  operator
$\Sigma_{e-e}(\omega)$. The latter is calculated in the framework
of the so-called 3BS formalism~\cite{cala-mang94prb,ThreeBS}. 3BS
treats the electron correlation in a full many body approach
beyond the single particle approximation. The resulting
self-energy is non-hermitian and frequency-dependent and the
inclusion of this quantity in the evaluation of quantum
transmittance may introduce incoherent and dynamical effects on
transport that are not accessible~\cite{ferr+05prb} within mean
field techniques, like {\em e.g} DFT+U.

The 3BS scheme requires the inclusion of an on-site Hubbard-like
term U for the Cu ions. We looked at the effects induced by two
values of the parameter, U=3.0 and U=6.0 eV. The results are
summarized in Fig.~\ref{figure3}(b). Similar to the single
molecule case, the effect of correlation does not change radically
the character of the up and down transmittance spectra. This is in
agreement with the absence of Cu$_{3d}$-derived states close to
the Fermi level. The displacement of the $b_{1g}$ state is not
sufficient to modify the conduction properties of the device. The
main differences are in the region $E\in [-2.0, -0.5]$ eV, where
an enhancement of the transmittance is observed. This is mainly
due to the real part of the e--e self-energy, which shifts the
molecular orbitals, leading a higher alignment with the lead
states. The amount of such realignment depends on the value of the
selected Hubbard U term, and it is not strictly related to a
further hybridization of the conductor/lead molecular orbitals.

%*******************************
%  MnPc
%*******************************

\subsection{Manganese(II)-phtalocyanine systems}\label{mnpc_mol}

\begin{figure}
\begin{center}
\includegraphics[width=1.0\textwidth]{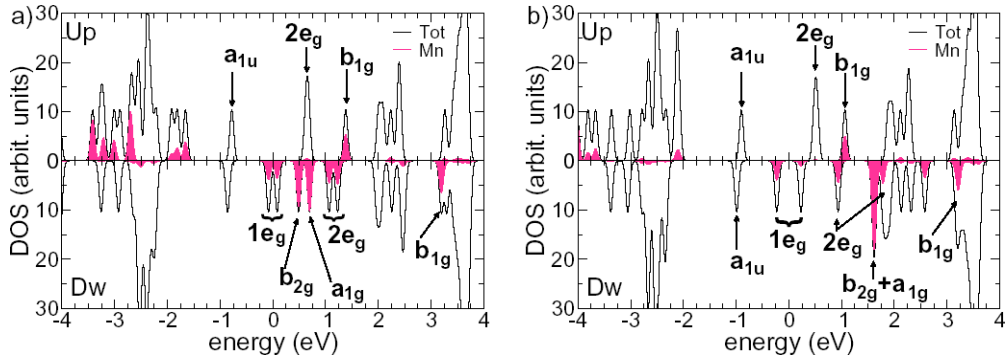}
\caption{(color online) MnPc electronic structure. Spin-resolved
density of states (DOS) corresponding to (a) PBE calculation  and
(b) PBE+U (U= 3eV) functionals. Shaded pink areas  are the
projections on the atomic Mn states. Vertical arrows and labels
identify the energy position of the peaks corresponding to
molecular orbitals described in text. }\label{figure4}
\end{center}
\end{figure}

\subsubsection{Single molecule.}
Let's now consider how the electronic structure of the molecule is
modified by the presence of a highly correlated Mn central cation
that substitute the original Copper. Due to the half occupation of
the strongly localized $3d$ shell, LDA is expected to fail in the
description of the electronic structure of the manganese atom.
Therefore, we followed previous theoretical calculations
\cite{liao+05ic} in using a gradient corrected (PBE) XC
functional.

The relaxed MnPc molecule still exhibits a planar square-like
geometry, as in the case of CuPc and we can still label  the
single molecular orbitals according to the $D_{4h}$ symmetry.
However, due to different ionic radius and different metal $d$
orbital occupancy, we observe a contraction ($\sim 1.5\%$) of the
Mn-N bond length and a reduction of the metal-ligand ionicity,
resulting in a stronger metal-molecule coupling. The spin-resolved
density of states of the isolated MnPc molecule is displayed in
Fig. \ref{figure4}(a). For the spin-up channel $a_{1u}$ and
$2e_{g}$ states are the single occupied HOMO$_{Up}$ and
LUMO$_{Up}$ respectively, while the $b_{1g}$ orbital is empty. The
energy difference $\Delta$E=E($2e_{g}$)-E($a_{1u})$=1.41 eV is
very similar to the CuPc case. On the contrary, the spin-down
$a_{1u}$ state correspond to the HOMO-1. The spin-down HOMO and
LUMO are given by the almost degenerate ($\Delta$E=0.16 eV) duplet
$e_{1g}$ (one state occupied, one empty). These are $\pi$-like
orbitals extended over both the Mn ion and the Pc rings. The
corresponding spin-up states are instead completely occupied,
where the double degenerate $2e_{g}$ peak is shifted at higher
energy and split into two single peaks, separated by
($\Delta$E=0.07 eV). Furthermore, they both exhibit a great
$3d$-Mn component. It is worth noticing that between the $e_{1g}$
and the $2e_{g}$ duplets there are two other empty states,
$b_{2g}$ and $a_{1g}$, fully occupied in the spin-up spectrum,
that are $\sigma$-like orbitals localized on the central ion.
Finally, the $b_{1g}$ peak is more than 3eV above the spin-down
HOMO. The MnPc molecule has a total magnetization of 3.0 $\mu_B$,
resulting from the unbalancing of three orbitals
($e_{1g}$,$b_{2g}$ and $a_{1g}$) between  the up and down
components.

The small SOMO-LUMO gap, along with the presence in the same
energy region of several molecular orbitals mostly localized on
the metallic site, deserves further investigation in view of the
strongly correlated character of the Manganese atom. To this end,
we improved the description of the  electronic structure by
including a Hubbard U term  on the central ion (DFT+U). We
considered two cases, U=2.0 eV and U=3.0 eV; the results for the
latter value are shown in Figure \ref{figure4}(b). Due to stronger
correlations effects, even the presence of small Hubbard potential
strongly modifies the electronic properties of the system. We
observe a general downward (upward) shift of the occupied (empty)
Mn components in the density of states. The effect of the
correlation is weaker for spin-up states, since the frontier
orbitals have a predominant Pc character. On the contrary, in the
spin down case the on-site e--e correlation increases the
SOMO-LUMO gap, splitting the almost degenerate $e_{1g}$ peak with
an energy separation that is a function of the applied U:
$\Delta$E=0.39 eV and $\Delta$E=0.46 eV for U=2.0 eV and U=3.0 eV
respectively. The $2e_{g}$ duplet also splits in two separates
contributes as shown in Figure \ref{figure4}(b). The effect of the
correlation on the $2e_{g}$ states is more relevant for the
minority spin, because it includes a greater metallic
contribution. The $b_{2g}$ and $a_{1g}$ peaks migrate towards
higher energies, between the split $2e_{g}$ fork, changing the
global order of the MnPc orbitals. Higher values of the local
Hubbard term (U$>3.0$ eV) do not lead to stable (converged)
electronic structures, confirming that the results are very
sensitive to the electron correlation model.

\subsubsection{Device configuration.}

\begin{figure}
\begin{center}

\includegraphics[width=1.0\textwidth]{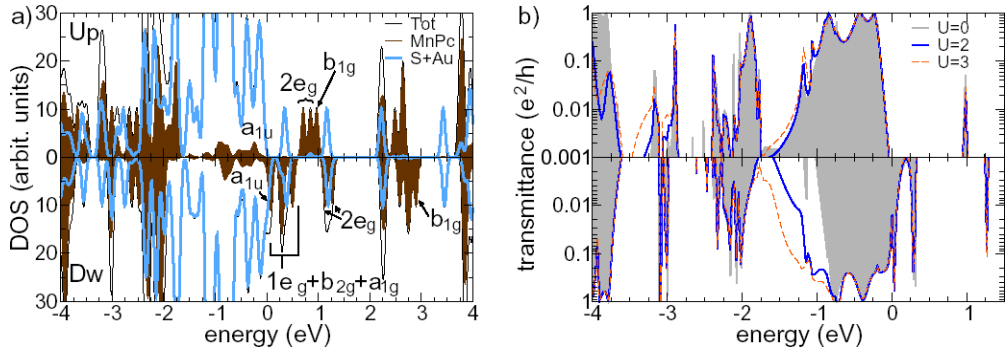}
\caption{(color online) (a) Spin-polarized DOS of periodic
(Au$_5$S)$_2$MnPc system. The shaded brown area (thick blue line)
is the projection on the MnPc (S-Au) atomic states. (b) Mean field
(shaded area) and correlated (straight and dashed lines)
two-terminal spin-resolved transmittance of an open MnPc-based
device. The zero energy references are set to the Fermi level of
the periodic (Au$_5$S)$_2$MnPc system.}\label{figure5}
\end{center}
\end{figure}

We modeled the double interface (Au$_5$S)$_2$MnPc as before in the
case of the CuPc. The relaxed geometry does not present  any new
characteristics, while the DFT electronic structure structure
[Fig. \ref{figure5}(a)] is more complex, due to the greater
complexity of the MnPc molecule. While the Au component of the
spectra is not particularly affected by the change of the metal
atom, the molecular states display a complex reorganization. In
the spin-up component, the molecule/metal coupling induces a
broadening of the $a_{1u}$ state with the $sd$ gold band in the
energy range E$\in[-1.0,-0.25]$ eV, and the splitting of the
pristine double degenerate $2e_{g}$ peak into a molecular doublet,
close to the empty $b_{1g}$ state. The spin-up HOMO and LUMO have
both  a $sd$ character, derived from the gold chain, and are
separated by an energy gap $\Delta$E=0.43 eV, higher than the
corresponding Cu case. The analysis of the single particle
orbitals shows that the spin-down HOMO has also a pure Au
character, while the spin-down LUMO is a hybrid state, stemming
from the mixing of the $sd$ Gold band and the $a_{1u}$ state of
the Phthalocyanine. The spin-down HOMO and LUMO are almost
degenerate in energy, with a gap $\Delta$E=74 meV [Fig.
\ref{figure5}(a)]. In the range E$\in[0.1,0.5]$ eV we observe a
manifold of states, which includes the two $e_{1g}$, the $b_{2g}$
and the $a_{1g}$ orbitals, each partially hybridized with the Gold
states. The split $2e_{g}$ and $b_{1g}$ peaks are recognizable at
higher energies. The depletion of the spin-down $a_{1u}$ and
$e_{1g}$ states, occupied in the gas phase (Fig. \ref{figure4}),
leads to a further unbalance of $\sim$2 electrons between the up
and down components, with respect to the isolated molecule. This
enhances the total magnetization up to $\mu$=4.79 Bohr mag/cell,
which may be justified in terms of charge transfer at the
interface from the molecule towards the gold chain for the
spin-down component.

This marked spin asymmetry is reflected in the transmittance
spectra, as shown in Fig.~\ref{figure5}(b). At the mean field
level, the spin-up component behaves as a semiconductor with a
vanishing conductance at zero bias. On the contrary, the
non-negligible density of states at the Fermi energy (zero of the
energy scale) in the spin-down component leads to a finite
conductance $G_{dw}=0.12$ e$^2/h$. If we compare the cases of Cu-
and Mn-Pc we observe a distinctively different behavior:  in the
case of Copper the reduced spin unbalance modifies only slightly
the almost degenerate conduction properties of the gold wires,
leading to a better conductor (higher conductance) for both the
spin channels, while for Manganese the spin asymmetry along with
an enhanced molecule/lead coupling changes the conduction
properties of the Au wire as a function of the spin. This is a
very remarkable result that suggests that  the MnPc molecular
device could work as a spin valve. If confirmed by experiments,
this would open the way to a new generation of devices for
molecular spintronics applications.

We have studied the correlated transmittance of the MnPc two
terminal device for two values of the Hubbard U parameter (U=2.0,
3.0 eV) in our 3BS scheme. The results for the
correlated transport [straight and dashed lines in
Fig.~\ref{figure5}(b)] clearly show that in the region near the
Fermi energy the transmittance is almost insensitive to the
inclusion of the electron-electron correlation. In particular,the
spin valve behavior of the MnPc-based device is stable upon the
inclusion of the e--e coupling, and only minor modifications at
E$\in[-2.0,-1.0]$ eV are induced by the realignment of the lower
energy molecular orbitals, as in the case of CuPc system.

This result can be understood in terms of the electronic structure
[Fig.~\ref{figure5}(a)]. The strong molecule/gold bonding leads to
a removal of the Mn$_{3d}$ states from the Fermi level to higher
energies, while the only molecular components close to $E=0$ eV
are related to the aromatic Pc rings ($a_{1u}$), only slightly
affected by the electron correlation. In other words, the short
range electron correlation truly affects transport when  the
simultaneous mixing of the delocalized states of the leads and the
partially occupied states of the correlated conductor occurs at
the Fermi energy \cite{ferr+05prl,ferr+05prb}. As a consequence,
both Cu- and Mn-Pc transmittances are almost unaffected by the
e--e correlation.

%*******************************

%  BIBLIOGRAPHY

   %*******************************

\section{Conclusions}

In conclusion, we have demonstrated that a proper inclusion of
electronic correlation effects is essential to develop a complete
understanding of the the electronic and transport properties of
phthalocyanine- based nanostructures. Considering different forms
for the exchange-correlation functional (LDA, PBE, DFT+U) in the
standard DFT, as well as the inclusion of on-site correlation
beyond DFT (3BS), we have proved that the inclusion of correlation
reproduces the correct orbital sequence in CuPc, as observed in
experiments. We should stress that our results are valid for the
molecule in its ground state, so that a comparison with
experiments involving oxidized or reduced states of the molecule
are not physically meaningful. Moreover, electronic correlation
have clearly an even greater effect in the MnPc molecule, due to
the open $3d$ shell of the Manganese ion. We do not exclude also
that other kind of effects, such as those related to the
Jahn-Teller distortions, might play an important role in the
electronic properties of these systems, as proposed in the case of
alkali doped  MPc crystals \cite{tosa+04prl}.

The transport properties of these
M(II)Pc-based devices are less sensitive to the inclusion of electron
correlation effects than the corresponding single molecule electronic
structures. The e-e interaction seems in fact to be partially
quenched by the strong coupling with the uncorrelated gold leads.
This observation confirms the fact that, due to the reduced number of
electronic states, the transport properties of a molecular device
can not be directly inferred from the electronic structure of the
isolated molecule. On the basis of our
transport results, we can not exclude {\em a priori}, that effects
of the correlation might be important in other device
configurations, where the molecule/lead coupling is different.

Finally, our electronic transport simulations have shown (irrespective on the details of the calculations) that
M(II)Pc-based devices can act selectively as molecular conductors,
as in the case of Copper, or as spin valves, as in the case of
Manganese, demonstrating once more the great potential of these
systems for molecular nanoelectronics applications.

\section*{Acknowledgments} Carlo Cavazzoni, Franca Manghi and Silvia Picozzi are
gratefully acknowledged for fruitful collaborations and
discussions. Funding was provided in part by the Regional
Laboratory of EmiliaRomagna ``Nanofaber''; by the Italian MIUR
through PRIN 2006, and by the Department of Energy of the US
government. Computing time at the CINECA supercomputing facilities
was provided by INFM-CNR.

\section*{References}

\bibliography{all}

\begin{thebibliography}{10}
\bibitem{note_hubbard}
The screened value of the local Hubbard potential may be evaluated in the self
  consistent calculation as the derivative of the total energy with respect to
  the number of electrons, as described in Ref. \cite{coco-degi05prb}. However
  this goes beyond the aim of the present work. Here, we are not interested in
  the exact evaluation of the e-e interaction, but in the qualitative
  description of the effects of the electron correlation.
\bibitem{note_hubbard1}
We observed this effect in this specific CuPc molecular structure, thus this
  conclusions can not be straightforwardly extended to other copper-based
  system, without deep further analysis.
\end{thebibliography}

\end{document}